\documentclass[aps,prd,twocolumn,groupedaddress,showpacs]{revtex4}

\usepackage{graphicx}
\usepackage{color}

\begin{document}


\title{General limit on the relation between abundances of D and $^7$Li in big bang nucleosynthesis with nucleon injections}


\author{Motohiko Kusakabe$^{1,2}$}
\email{motohiko@kau.ac.kr} 
\author{Myung-Ki Cheoun$^{2}$}
\email{cheoun@ssu.ac.kr} 
\author{K. S. Kim$^{1}$}
\email{kyungsik@kau.ac.kr} 

\affiliation{
$^1$School of Liberal Arts and Science, Korea Aerospace University, Goyang 412-791, Korea}
\affiliation{
$^2$Department of Physics, Soongsil University, Seoul 156-743, Korea}


\date{\today}

\begin{abstract}
The injections of energetic hadrons could have occurred in the early universe by decays of hypothetical long-lived exotic particles.  The injections induce the showers of nonthermal hadrons via nuclear scattering.  Neutrons generated at these events can react with $^7$Be nuclei and reduce $^7$Be abundance solving a problem of the primordial $^7$Li abundance.  We suggest that thermal neutron injection is a way to derive a model independent conservative limit on the relation between abundances of D and $^7$Li in a hadronic energy injection model.  We emphasize that an uncertainty in cross sections of inelastic $n+p$ scattering affects the total number of induced neutrons, which determines final abundances of D and $^7$Li.  In addition, the annihilations of antinucleons with $^4$He result in higher D abundance and trigger nonthermal $^6$Li production.  It is concluded that a reduction of $^7$Li abundance from a value in the standard big bang nucleosynthesis (BBN) model down to an observational two $\sigma$ upper limit is necessarily accompanied by an undesirable increase of D abundance up to at least an observational 12 $\sigma$ upper limit from observations of quasi-stellar object absorption line systems.  The effects of antinucleons and secondary particles produced in the hadronic showers always lead to a severer constraint.  The BBN models involving any injections of extra neutrons are thus unlikely to reproduce a small $^7$Li abundance consistent with observations.
\end{abstract}



\maketitle

\section{Introduction}\label{sec1}

Many environments have been considered regarding the origin of deuterium~\cite{Epstein1976Natur.263..198E}.  They include pregalactic cosmic rays (CRs) from quasars and collapsing objects, shock waves, and neutron stars.  In general, the CRs induce nuclear reactions producing D, $^3$He, Li, Be, and B nuclides~\cite{Reeves1970Natur.226..727R,Meneguzzi1971A&A....15..337M,Reeves1974ARA&A..12..437R}.  Pregalactic CRs or cosmological CRs generated before the Galaxy formation also produce $^{6,7}$Li (via the $\alpha+\alpha$ fusion~\cite{Montmerle1977ApJ...216..177M}) and $^3$He (via $^4$He+$p$ nuclear spallation~\cite{Montmerle1977ApJ...217..878M}).  $^6$Li productions have been calculated for the CRs in specific environments:  the CRs accelerated in structure formation shocks at the Galaxy formation epoch~\cite{2002ApJ...573..168S} and the CRs from supernova remnants at the pregalactic epoch~\cite{Rollinde:2004kz,Rollinde:2006zx}.  Since a metal pollution proceeds along with a stellar activity in the universe, the CRs would come to contain metals such as C, N, and O.  Therefore, the pregalactic CR nucleosynthesis would also produce Be and B through reactions of (C, N, or O)+($p$ or $\alpha$)~\cite{Kusakabe2008,Rollinde2008} and ($^3$He or $\alpha$)+$\alpha\rightarrow (^6$He or $^{6,7}$Li)+$a$ followed by ($^6$He or $^{6,7}$Li)+$\alpha\rightarrow ^9$Be+$b$ with byproducts $a$ and $b$~\cite{Kusakabe:2012sh}.

Another possible source of the CR is an energy injection at decay and annihilation of exotic long-lived particles~\cite{Lindley1979MNRAS.188P..15L,1982NCimR...5j...1C,Khlopov:1984pf,Balestra:1984cu,Ellis:1984er,Lindley:1986wt,Sedelnikov:1987ef,Reno:1987qw,Levitan:1988au,Dimopoulos:1987fz,Dimopoulos:1988zz,Dimopoulos:1988ue,Terasawa:1988my,Kawasaki:1993gz,Khlopov:1993ye,Kawasaki:1994af,Holtmann:1996cq,Jedamzik:1999di,Kawasaki:2000qr,Cyburt:2002uv,Kawasaki:2004yh,Kawasaki:2004qu,Jedamzik:2004er,Jedamzik:2004ip,Jedamzik:2005dh,Ellis:2005ii,Kusakabe:2006hc,Kanzaki:2006hm,Jedamzik:2006xz,Cumberbatch:2007me,Kusakabe:2008kf,Kawasaki:2008qe,Kawasaki:2009ex,Cyburt:2009pg,Pospelov:2010cw,Pospelov:2010kq,Cyburt:2010vz,Ellis:2011sv,Kang:2011vz,Olive:2012xf,Kusakabe:2013sna,Ishida:2014wqa}.  A constraint on the mass of a hypothetical stable heavy neutrino has been derived through calculation of its present cosmological energy density~\cite{Hut:1977zn,Lee:1977ua}.  An unstable heavy neutrino was then considered, and constraints on its mass and lifetime were derived~\cite{Sato:1977ye,Dicus:1977nn,Vysotsky:1977pe}.  The electromagnetic decay of the unstable particle is constrained through distortions in the energy spectrum of cosmic microwave background radiation~\cite{Sato:1977ye}.  The constraints on hypothetical heavy neutrino~\cite{Miyama:1978mn} and primordial black holes~\cite{Miyama:1978mp} were then derived from the effect on light element abundances through energy densities in detailed calculations of big bang nucleosynthesis (BBN).  The decay of unstable heavy neutrinos also affects nuclear abundances through nonthermal photodissociation of nuclei~\cite{Lindley1979MNRAS.188P..15L}.  The radiative decay induces electromagnetic cascades of energetic photons, electrons, and positrons during the propagation of the nonthermal photon emitted at the decay~\cite{Ellis:1984er}.

Effects of hadronic injections at the decay were studied~\cite{Levitan:1988au,Dimopoulos:1987fz,Dimopoulos:1988ue,Dimopoulos:1988zz,Reno:1987qw}.  Levitan {\it et al.} investigated hadronic cascades of proton and antiproton and dissociations of $^4$He~\cite{Levitan:1988au}.  Dimopoulos {\it et al.}~\cite{Dimopoulos:1987fz,Dimopoulos:1988ue,Dimopoulos:1988zz} extensively studied the effects on abundances of nuclei up to $^7$Li and $^7$Be.  They considered the reaction, i.e., $^1$H($n,\gamma$)$^2$H, for D production, and the reaction, i.e., $^7$Be($n,p$)$^7$Li, for $^7$Be destruction, where 1(2$,$3)4 stands for a reaction $1+2\rightarrow 3+4$.  Antiprotons injected at decays of exotic long-lived particles could dissociate $^4$He and produce D and $^3$He~\cite{1982NCimR...5j...1C,Khlopov:1984pf,Balestra:1984cu}.  The cross sections of $\bar{p}+^4$He annihilation have been measured~\cite{1988NCimA.100..323B}, and the yields of D, $^3$H, and $^3$He at the annihilation were calculated as a function of energy of antiproton~\cite{Sedelnikov:1999km}.  Effects of exotic particles on nuclear abundances through hadronic showers have been extensively studied with realistic initial spectra of injected hadrons~\cite{Kawasaki:2004qu,Jedamzik:2006xz}.

The standard BBN (SBBN) model explains primordial light element abundances inferred from astronomical observations well~\cite{2011ARNPS..61...47F}.  Modifications of the BBN model are then constrained from the consistency between theoretical predictions and observations of abundances.  Among light elements produced during the BBN, however, the lithium has an unexplained discrepancy between SBBN prediction and observational determinations of its primordial abundances~\cite{Melendez:2004ni,Asplund:2005yt}.   Spectroscopic observations of metal-poor stars (MPSs) indicate an abundance measured by number relative to hydrogen, i.e., $^7$Li/H$=(1-2) \times 10^{-10}$~\cite{Spite:1982dd,Ryan2000,Melendez:2004ni,Asplund:2005yt,bon2007,Shi:2006zz,Aoki:2009ce,Hernandez:2009gn,Sbordone2010,Monaco:2010mm,Monaco:2011sd,Mucciarelli:2011ts} \footnote{Surface Li abundances of metal-poor red giant branch stars do not depend on parameters of standard stellar models as much as dwarf stars do.  Mucciarelli {\it et al.}~\cite{Mucciarelli:2011ts} determined Li abundances of metal-poor halo red giant branch stars, and estimated initial abundances, which were also $\sim$2--3 lower than SBBN prediction.} \footnote{Monaco {\it et al.}~\cite{Monaco:2011sd} reported that one star, $\sharp$37934, among 91 stars of the globular cluster M4 has a high lithium abundance ($^7$Li/H=$7.4^{+3.1}_{-2.2}\times 10^{-10}$) consistent with the abundance of the SBBN model.}.  This abundance is a factor of 2--4 higher than the SBBN prediction when we adopt the baryon-to-photon ratio determined from the observation of the cosmic microwave background radiation with Wilkinson Microwave Anisotropy Probe (WMAP)~\cite{Hinshaw:2012aka}.

After the lithium problem was recognized, the neutron injection during the BBN was suggested to be a solution since it can reduce $^7$Be abundance via $^7$Be($n,p$)$^7$Li($p,\alpha$)$^4$He, although it increases D abundance via $^1$H($n,\gamma$)$^2$H simultaneously~\cite{Jedamzik:2004er,Vasquez:2012dz}.  Such a neutron injection is realized in the hadronic decay of exotic long-lived massive particles~\cite{Jedamzik:2004er,Kawasaki:2004qu,Jedamzik:2006xz}.  Important reactions caused by injected nonthermal hadrons have been identified in a statistical study, which are shown to be closely associated with resulting elemental abundances~\cite{Cyburt:2010vz}.  A wide parameter region of the lifetime and the abundance of a long-lived particle was studied, and a parameter region for $^7$Li reduction has been found~\cite{Kawasaki:2004qu,Jedamzik:2006xz,Cyburt:2009pg} \footnote{If long-lived exotic particles of sub GeV-scale mass exist, and their decay products do not include nucleons, another route of additional neutrons operates~\cite{Pospelov:2010cw}.  When mesons such as $\pi$ and $K$ are generated by the particle decays, they can convert protons to neutrons, and a reduction of $^7$Li abundance realizes along with an enhancement of D abundance.  When the decays do not generate any mesons, and muons and neutrinos are generated, on the other hand, induced electron antineutrinos convert protons to neutrons.  In this case, the dissociation of once enhanced D by nonthermal photons can reduce D abundance to the level consistent with observations.}.  In this paper, we focus solely on the parameter region for $^7$Li reduction, and derive a model independent constraint on a relation between abundances of D and $^7$Li, by using recent D abundance data.

In Sec.~\ref{sec2}, we describe input physics and assumptions adopted in this paper.  We prove that the assumption of thermal neutron injection (TNI) leads to a conservative lower limit on the ratio of the increase of D abundance to the decrease of $^7$Li abundance.  In Sec.~\ref{sec3}, we describe the TNI model and the BBN model, as well as adopted observational constraint on primordial nuclear abundances.  The TNI is assumed to occur instantaneously, and the injection time and the abundance of injected neutron are used as parameters in this model.  In Sec.~\ref{sec4}, results of the BBN calculations are shown, and a relation between abundances of D and $^7$Li is derived.  In Sec.~\ref{sec5}, we estimate an effect of antinucleon annihilation with $^4$He on the abundance relation.  In Sec.~\ref{sec6}, we estimate amounts of $^6$Li production induced by the antinucleon+$^4$He annihilation.  In Sec.~\ref{sec7}, conclusions are done finally.  In Appendix~\ref{app1}, we list important nuclear reactions which work in a parameter region for the reduction of primordial $^7$Li abundance.  In Appendix~\ref{app2}, approximate analytic estimates of D and $^7$Li abundances are shown.

In this paper, we adopt notation of $a(n)=a\times 10^n$ with a real number $a$ and an integer $n$, and $Q_{,b}=Q/b$ with a parameter $Q$ and a real number $b$.  The Boltzmann's constant ($k_{\rm B}$), the reduced Planck's constant ($\hbar$), and the light speed ($c$) are normalized to be unity.

\section{input physics}\label{sec2}

In this paper, we concentrate on a production of D and a reduction of $^7$Be and $^7$Li induced by hadronic energy injection at temperature $T_9\equiv T/(10^9~{\rm K})\gtrsim 0.4$ (or cosmic time $t\lesssim 10^3$~s).  This injection epoch corresponds to that of the solution to the $^7$Li problem by the hadronic energy injection model~\cite{Jedamzik:2004er,Jedamzik:2004ip,Kawasaki:2004qu}.  The injection produces energetic nucleons, antinucleons, and mesons.  Such hadrons can scatter background nuclei so that many energetic hadrons are generated and hadronic showers composed of energetic hadrons are developed~\cite{Dimopoulos:1987fz,Dimopoulos:1988ue,Kawasaki:2004qu,Jedamzik:2006xz}.  Main reactions changing abundances of D and ($^7$Li+$^7$Be)~\footnote{A sum of $^7$Li and $^7$Be abundances gives primordial $^7$Li abundance before the start of the early stellar activity.  This is because $^7$Be produced at BBN epoch is transformed to $^7$Li by electron capture after the recombination of $^7$Be$_{\rm V}$ ion.} in this parameter range ~\cite{Jedamzik:2004er,Kawasaki:2004qu,Jedamzik:2006xz} are
\begin{eqnarray}
^1{\rm H}(n,\gamma)^2{\rm H}~~~{\rm and}~~~
^4{\rm He}(n,x)^2{\rm H},
\label{eq1}
\end{eqnarray}
\begin{equation}
^7{\rm Be}(n,p)^7{\rm Li}(p,\alpha)^4{\rm He},
\label{eq2}
\end{equation}
respectively,
where
$x(=^3$H, $d+n$, or $p+2n)$ is a byproduct.  If the neutron injection time is $t_{\rm inj}\sim 10^3$~s as considered here, effects of long-lived mesons are negligible \cite{Reno:1987qw}.

In Sec.~\ref{sec21}, we comment that energetic proton, antiproton, and nuclei quickly thermalize while an energetic neutron can induce inelastic scatterings off background proton.  In Sec.~\ref{sec22}, we present that the BBN calculation for the case of the TNI provides a lower limit on the ratio, i.e., $\Delta$D/$|\Delta^7$Li$|$, where $\Delta A\equiv n_A-n_{A,{\rm SBBN}}$ is a difference between final number densities of nuclide $A$ in this model ($n_A$) and the SBBN model ($n_{A,{\rm SBBN}}$).  A more precise estimation of the ratio $\Delta$D/$|\Delta^7$Li$|$ should include the annihilation of antineutron with $^4$He.  Nuclear data on the annihilation, however, contains a large uncertainty.  It is shown that effects of hadronic showers composed of energetic neutron and antineutron always enhance the ratio $\Delta$D/$|\Delta^7$Li$|$ above that of the TNI model.  In Sec.~\ref{sec23}, we see that the assumption of instantaneous thermalization of nonthermal neutron leads to a lower limit on the ratio.

\subsection{Hadronic shower}\label{sec21}
\subsubsection{Stopping of energetic proton}
We assume instantaneous thermalizations of nonthermal $p$, $\bar{p}$, and nuclei for the following reason.

An inelastic scattering of two nucleons can be triggered by incident nucleons with energies of $\gtrsim {\mathcal O}(0.1)$~GeV (Fig. 1 in Ref.~\cite{Meyer1972A&AS....7..417M}).  Such incident nucleons are thus relativistic to a certain degree.  Because of the Coulomb interaction via electric charge, a relativistic proton undergoes Coulomb energy loss.  The loss rate for $T<m_e$ is given (Eq. [A.18] in Ref.~\cite{Reno:1987qw} \footnote{We checked this equation.  The Eq. (B6) in Ref.~\cite{Kawasaki:2004qu} may show an erroneously larger rate by a factor of two.}) by
\begin{equation}
 \frac{dE}{dt}=-\frac{2\pi Z^2\alpha^2\Lambda}{m_e^2}\rho_e,
\label{eq3}
\end{equation}
where
$E$ and $Z=1$ is the kinetic energy and the charge number of proton, respectively,
$\alpha$ is the fine-structure constant,
and $m_e$ is the electron mass.
$\Lambda\approx\ln(m_ev^2\gamma^2/\omega_{\rm p})$ is a parameter associated with Coulomb divergence (Eqs. [13.13] and [13.43] in Ref.~\cite{1975clel.book.....J}).  Here, $v$ is the velocity of the proton, and $\gamma=(1-v^2)^{-1/2}$ is the Lorentz factor.  $\omega_{\rm p}=\sqrt[]{\mathstrut 4\pi\alpha n_e/m_e}$ is the plasma frequency of background plasma composed of electron and positron~\cite{1975clel.book.....J}.  $n_e$ and $\rho_e\sim m_e n_e$ are the total number density and the energy density, respectively, of electron and positron plasma.  The total number density is given by $n_e\sim 4[m_eT/(2\pi)]^{3/2}\exp(-m_e/T)$ for $m_e > T\geq m_e/26$ and $n_e\sim (1-Y/2)\eta n_\gamma$ for $T< m_e/26$ with the mass fraction $Y$ of $^4$He to total baryon, the baryon-to-photon number ratio $\eta$, and the number density $n_\gamma$ of background photon~\cite{Reno:1987qw}.

Cross sections for inelastic scattering of two nucleons are $\sigma\lesssim 30$ mb \cite{Meyer1972A&AS....7..417M}.  The reaction rate is then given by
\begin{eqnarray}
 \Gamma_{\rm inel}&\sim& n_p \sigma v\nonumber\\
&=& 5.4\times 10^2~{\rm s}^{-1}~T_{9, 0.4}^3~\eta_{,6.2(-10)}~X_{,0.75}~\sigma_{,30~{\rm mb}}~v. \nonumber\\
\label{eq4}
\end{eqnarray}
The rate of energy degradation via Coulomb scattering is, on the other hand, given by
\begin{eqnarray}
 \Gamma_{\rm loss}&\sim& \frac{1}{E}\left|\frac{dE}{dt}\right|=\frac{2^{3/2}Z^2\alpha^2 \Lambda m_e^{1/2} T^{3/2}}{\pi^{1/2} E}\exp(-m_e/T)\nonumber\\
&=&1.4\times 10^6~{\rm s}^{-1}~Z^2 T_{9, 0.4}^{3/2}~\Lambda_{,6.4}~\left[\frac{{\mathrm e}^{-m_e/T}}{3.6(-7)}\right]~E_{,{\rm GeV}}^{-1},\nonumber\\
\label{eq5}
\end{eqnarray}
where
the numerical factor in the second line corresponds to the case of $T_9=0.4$.
Nonthermal protons generated at hadron injections hardly trigger an inelastic collision before they lose energies because of quick thermalization, i.e., $\Gamma_{\rm loss}\gg \Gamma_{\rm inel}$ for $T_9>0.4$.  The same holds true for antiprotons, and nuclei with larger charge numbers.

\subsubsection{Inelastic scattering of energetic neutron}

Nonthermal protons effectively stop without inducing hadronic scatterings.  Hadronic showers then contain only neutrons and antineutrons as mediator particles which can interact with background nuclei nonthermally by the energy injected at the particle decay.  Main reactions between a nonthermal neutron and a background proton, which is much more abundant than background neutron at $T_9\lesssim 0.4$, are
\begin{eqnarray}
 n+p&\rightarrow&n+p+N\pi^0+M(\pi^+\pi^-)
\label{eq6}\\
n+p&\rightarrow&n+n+\pi^++N\pi^0+M(\pi^+\pi^-)
\label{eq7}\\
n+p&\rightarrow&p+p+\pi^-+N\pi^0+M(\pi^+\pi^-),
\label{eq8}
\end{eqnarray}
where
$N$ and $M$ are nonnegative integers.
The elastic scattering corresponds to $N=M=0$ in Eq. (\ref{eq6}).  For a same set of $N$ and $M$ values, reaction thresholds of the second reaction are higher than those of the third by $2(m_n-m_p)$, where $m_n$ and $m_p$ are the masses of neutron and proton, respectively.

The first reaction does not change the combination of nucleon isospins so that the number of energetic particle, i.e., neutron, is not changed.  The second reaction could increase the number of energetic neutron, while the third decreases it both by the unit of one.  Two protons from the third reaction stop instantaneously.  If the sum of rates for the second reaction over $N$ and $M$ is larger than that for the third, nonthermal neutron abundance goes up from the abundance of originally injected neutron.  If the total rate for the second is smaller than that for the third, however, the nonthermal neutron abundance goes down.  If the both rates balance approximately, the nonthermal neutron abundance does not change during developments of hadronic showers.

Cross sections of the second and third reactions have been measured, and they equate within the statistical errors~\cite{Dunaitsev1960, Bystricky1987}.  Although an isospin symmetry in the two reactions seems to exist, it is not yet verified experimentally.  Uncertainties in reaction rates affect a net number of neutrons which are generated in the universe.  The net abundance of nonthermal neutron is the most important quantity determining abundances of D and $^7$Li.  Then, one should be cautious about the uncertainties in reaction rates when a parameter space for $^7$Li reduction is searched.  Recent previous BBN calculations including hadronic particle injection were based on biased network codes in which either reaction of the second and third types is included for some sets of $N$ and $M$~\cite{Kawasaki:2004qu,Jedamzik:2006xz}.  The present study escapes from these uncertainties, and obtains a conservative lower limit on $\Delta$D/$|\Delta^7$Li$|$.

\subsection{Production of neutron and D}\label{sec22}

In this subsection, we focus on the processes occurring at the time of neutron injection, $t_{\rm inj}$, and omit the index for the time $t_{\rm inj}$ on physical quantities for simplicity.  Firstly we describe changes in D and $^7$Li abundances caused by injections of neutrons and antineutrons by the following two equations.  The amount of $^7$Li reduction is approximately proportional to the total abundances of injected nonthermal neutron, i.e., $\Delta n_{\rm inj}~(>0)$ since $^7$Be is destroyed by neutron [Eq. (\ref{eq2})].  The equation for $\Delta n_{\rm inj}$ is
\begin{eqnarray}
\Delta n_{\rm inj} &=& n_1\left\{1+P_{1\rightarrow 2}(n)\left[1+P_{2\rightarrow 3}(n)(1+\cdots)\right]\right\}\nonumber\\
&&+\bar{n}_1 P^\prime_{1\rightarrow 2}(n)\left[1+P_{2\rightarrow 3}(n)(1+\cdots)\right],
\label{eq9}
\end{eqnarray}
where
$n_1$ and $\bar{n}_1$ are the abundances of primary neutron and antineutron, respectively, injected at the considered event,
$P_{N\rightarrow N+1}(i)$ and $P^\prime_{N\rightarrow N+1}(i)$ are the probabilities that the $N$-th generation neutrons and antineutrons, respectively, generate the $N+1$-th generation species for $i=n$ or $d$.  We note that $P_{N\rightarrow N+1}(n) = \sum_{j} P^j_{N\rightarrow N+1}(n)$ is a sum of components for multiple reactions ($j$).  If no neutron is emitted at a reaction induced by a $N$-th neutron, the net number of neutron changes by $-1$.  The $P^j_{N\rightarrow N+1}(n)$ value is then $-1$ for this reaction $j$.  The first and second terms of the right hand side (RHS) correspond to neutrons originating from primary neutrons and antineutrons, respectively.  We neglect effects of the $\bar{n}$ scattering off background $p$ and $^4$He.  Since annihilation cross sections of $\bar{n}p$ and $\bar{n}^4$He reactions are significant in comparison with total cross sections~\cite{1994NCimR..17f...1B}, generated $\bar{n}$ are typically lost after at most a few reactions unaccompanied with annihilations.

The change in D abundance is described as
\begin{eqnarray}
 \Delta {\rm D}&\hspace{-5pt}=&\hspace{-5pt}\Delta n_{\rm inj}\nonumber\\
&\hspace{-5pt}&\hspace{-5pt}+n_1 \left\{P_{1\rightarrow 2}(d)+P_{1\rightarrow 2}(n) \left[P_{2\rightarrow 3}(d)+P_{2\rightarrow 3}(n) \cdots\right]\right\}\nonumber\\
&\hspace{-5pt}&\hspace{-5pt}+\bar{n}_1 \left\{P^\prime_{1\rightarrow 2}(d)+P^\prime_{1\rightarrow 2}(n) \left[P_{2\rightarrow 3}(d)+P_{2\rightarrow 3}(n) \cdots\right]\right\},\nonumber\\
\label{eq10}
\end{eqnarray}
The first term of the RHS is for deuterons produced via $^1$H($n,\gamma$)$^2$H.  Note that the injected neutrons are mostly captured by proton, and converted to D for $t_{\rm inj}\sim 10^3$ s.  The second term is for the sum of the $N+1$-th deuterons produced mainly via $^4$He spallation by the $N(\geq 1)$-th neutrons which originate from primary neutrons.  The third term includes deuterons produced at annihilations with $^4$He, and the sum of the $N+1$-th deuterons produced mainly via $^4$He spallation by the $N(\geq 2)$-th generation neutrons originating from primary antineutrons.

The present model is constrained by an overproduction of D as described below.  We then conserve the model by keeping D abundances low while reducing $^7$Li abundances.  When instantaneous thermalizations of energetic $n$ and $\bar{n}$ are assumed, no secondary or higher order energetic particles would be generated.  Then, an equation, i.e., $P_{N\rightarrow N+1}(i)=0$, holds.  Accordingly, one obtains $\Delta n_1=n_1+\bar{n}_1 P^\prime_{1\rightarrow 2}(n)$, and $\Delta {\rm D}_1=\Delta n_1+\bar{n}_1 P^\prime_{1\rightarrow 2}(d)$, where subscript 1 in $\Delta n_1$ and $\Delta {\rm D}_1$ indicates that the amounts count only particles originating from primary neutrons and not higher order neutrons.

The $\Delta {\rm D}/\Delta n_{\rm inj}$ ratio is estimated as follows: First, we assume the symmetry in injected amounts of neutron and antineutron ($n_1=\bar{n}_1$).  The following relation then holds:
\begin{equation}
 \frac{\Delta {\rm D}_1}{\Delta n_1}=\frac{1+P^\prime_{1\rightarrow 2}(n)+P^\prime_{1\rightarrow 2}(d)}{1+P^\prime_{1\rightarrow 2}(n)}.
\label{eq11}
\end{equation}
The $P^\prime_{1\rightarrow 2}(i)$ value is given by
\begin{equation}
 P^\prime_{1\rightarrow 2}(i)=\left(\frac{n_\alpha\sigma_\alpha}{n_{\rm H}\sigma_p+n_\alpha\sigma_\alpha}\right)_{\bar{n}}~P_i(\bar{n}),
\label{eq12}
\end{equation}
where
$n_{\rm H}$ and $n_\alpha$ are number densities of $^1$H and $^4$He, respectively.  In the epoch after the $^4$He production, the ratio is $n_\alpha/n_{\rm H}=0.082$.  $\sigma_p$ and $\sigma_\alpha$ are cross sections for annihilation by hydrogen and $\alpha$ particle, respectively.  The ratio in the parenthesis with subscript $\bar{n}$ indicates the value for annihilation of $\bar{n}$.  $P_i(\bar{n})$ is the fraction of the $\bar{n}+^4$He annihilation into exit channels including species $i$.

\subsubsection{Effect of $\bar{n}$ annihilation}
Although an estimation of $P^\prime_{1\rightarrow 2}(i)$ [Eq. (\ref{eq12})] is associated with uncertainties, an example estimation is shown as follows:

Nuclear data on $\bar{p}+^4$He annihilation at low energies indicate fractions for the production of $d$ and $n$, i.e., $P_d(\bar{p})=0.07$--$0.18$ and $P_n(\bar{p})\lesssim 1-[P_{^3{\rm H}}(\bar{p})+P_{^3{\rm He}}(\bar{p})]<0.4$ \cite{1988NCimA.100..323B}.  We then assume the similarity of the fractions for $\bar{p}$ and $\bar{n}$, and take values of $P_d(\bar{n})=0.1$ and $P_n(\bar{n})\lesssim 0.4$.  In addition, we assume the simple scaling of $\sigma\propto A^{2/3}$ with the mass number $A$, and $\sigma_\alpha/\sigma_p=4^{2/3}$~\cite{Levitan:1988au}.  In this case, the equation, $P^\prime_{1\rightarrow 2}(i)=0.17~P_i(\bar{n})$, holds, and Eq. (\ref{eq11}) becomes
\begin{equation}
 \frac{\Delta {\rm D}_1}{\Delta n_1}\gtrsim 1.016.
\label{eq13}
\end{equation}

\subsubsection{Effect of secondary neutron}
Here the assumption of instantaneous thermalization is removed, i.e., $P_{N\rightarrow N+1}(i)\neq 0$.  A relation between yields of the $N(\geq 2)$-th generation neutron and deuteron derives from Eqs. (\ref{eq9}) and (\ref{eq10}) as
\begin{equation}
 \frac{\Delta {\rm D}_N}{\Delta n_N}=1+\frac{P_{N-1\rightarrow N}(d)}{P_{N-1\rightarrow N}(n)}.
\label{eq14}
\end{equation}
The quantity $P_{N-1\rightarrow N}(i)$ is described by an integration of a distribution function in energy of the $(N-1)$-th generation neutron multiplied by a rate for production of species $i$.  A lower limit on $P_{N-1\rightarrow N}(d)/P_{N-1\rightarrow N}(n)$ is estimated utilizing experimental data on cross sections~\cite{Meyer1972A&AS....7..417M} as
\begin{eqnarray}
 \frac{P_{N-1\rightarrow N}(d)}{P_{N-1\rightarrow N}(n)}&\sim& \frac{n_\alpha \sigma(n+\alpha\rightarrow d)}{n_p \sigma(n+p\rightarrow n) + n_\alpha \sigma(n+\alpha\rightarrow n)}\nonumber \\
&>&0.074,
\label{eq15}
\end{eqnarray}
where
$\sigma(n+i\rightarrow j)$ for $i=p$ and $\alpha$, and $j=d$ and $n$ represents an effective cross section for production of $j$ at the reaction with $i$, as explained below.
We defined
\begin{eqnarray}
\sigma(n+\alpha\rightarrow d)&=&\sigma(n+\alpha\rightarrow d+^3{\rm H}) \nonumber\\
&&+\sigma(n+\alpha\rightarrow d+p+2n) \nonumber\\
&&+2\sigma(n+\alpha\rightarrow 2d+n),
\label{eqadd1}
\end{eqnarray}
which is a sum of cross sections $\sigma(n+\alpha\rightarrow A_{\rm c})$ for final states $A_{\rm c}$ weighted according to the net increase in deuteron number.
Similarly we defined
\begin{equation}
\sigma(n+p\rightarrow n)=\sigma(n+p\rightarrow 2n+{\rm any})-\sigma(n+p\rightarrow 2p+{\rm any}),
\label{eqadd2}
\end{equation}
and
\begin{eqnarray}
\sigma(n+\alpha\rightarrow n)&=&\sigma(n+\alpha\rightarrow ^3{\rm He}+2n) \nonumber\\
&&+\sigma(n+\alpha\rightarrow d+p+2n) \nonumber\\
&&+2\sigma(n+\alpha\rightarrow 2p+3n) \nonumber\\
&&-\sigma(n+\alpha\rightarrow ^3{\rm H}+d),
\label{eqadd3}
\end{eqnarray}
as the sums of cross sections weighted according to the net increase in neutron number.

The value in the second line of Eq. (\ref{eq15}) was estimated as follows:  We adopt values of $\sigma(n+\alpha\rightarrow d)\gtrsim 10$~mb from the mirror reaction, i.e., $\sigma(p+\alpha\rightarrow d)$ (Fig. 7 of Ref.~\cite{Meyer1972A&AS....7..417M}), and $\sigma(n+\alpha\rightarrow n)\lesssim 60$~mb (Fig. 6 of Ref.~\cite{Meyer1972A&AS....7..417M}).  In addition, an asymmetry in cross sections of
$n+p\rightarrow 2n+{\rm any}$ [Eq.(\ref{eq7})] and $n+p\rightarrow 2p+{\rm any}$ [Eq.(\ref{eq8})] was allowed conservatively by 20 \% of the total inelastic cross section at maximum, i.e., $\sigma(n+p\rightarrow n)<\sigma(n+p)_{\rm inel}/5\sim 30/5$~mb~\cite{Meyer1972A&AS....7..417M}.

By comparing Eq. (\ref{eq13}) with Eqs. (\ref{eq14}) and (\ref{eq15}), it is found that an addition of contribution from the $N(\geq 2)$-th generation neutron always enhances the $\Delta {\rm D}/\Delta n_{\rm inj}$ ratio.

\subsection{Neutron thermalization}\label{sec23}
In this model, in addition to the SBBN, we consider an extra production of D and a destruction followed by some degrees of reproduction of $^7$Be (Sec.~\ref{sec4}).

We write the ratio between changes of ($^7$Li+$^7$Be) and D as a function of the kinetic energy of neutron, i.e., $E$, and $T$.  It is given by
\begin{equation}
\frac{|\Delta ^7{\rm Li}(E,T)|}{\Delta {\rm D}(E,T)}=\frac{n_{^7{\rm Be}}~\sigma(^7{\rm Be}+n,E)~P_{\rm des}(^7{\rm Li})}{n_{\rm H}~\sigma (^1{\rm H}+n,E)~P_{\rm sur}(D)},
\label{eq16}
\end{equation}
where
$\Delta A(E,T)$ is the change of $A$ abundance caused by neutrons with energy $E$ in the universe of temperature $T$.  $\sigma(A+n,E)$ is the cross section for the reaction $A+n$ as a function of $E$.

$P_{\rm des}(^7{\rm Li})$ is the destruction fraction of $^7$Li, which is produced via the reaction $^7$Be($n,p$)$^7$Li, during its propagation in the cooling universe.  $P_{\rm sur}({\rm D})$ is the survival fraction of D, which is produced via the reaction $^1$H($n,\gamma$)$^2$H, during its propagation.  If energetic $^7$Li and $d$ nuclei are produced by the respective reactions, they instantaneously lose their energies through the Coulomb scattering, and are thermalized soon after the productions~\cite{Cyburt:2002uv,Kawasaki:2004qu}.  The quantities $P_{\rm des}(^7{\rm Li})$ and $P_{\rm sur}({\rm D})$ should then be taken as values for thermal Maxwell-Boltzmann distribution of $^7$Li and D (see Appendix~\ref{app2}).  Note that although the quantities, $P_{\rm des}(^7{\rm Li})$ and $P_{\rm sur}({\rm D})$, depend on $T$, we omit to express the argument.

The ratio of $\sigma(^7{\rm Be}+n,E)/\sigma(^1{\rm H}+n,E)$ is roughly speaking smaller at higher energies as seen hereinbelow while the ratio of $P_{\rm des}(^7{\rm Li})/P_{\rm sur}({\rm D})$ is larger at higher energies (see Fig.~\ref{fig6} in Appendix~\ref{app2}).

Figure \ref{fig1} shows the ratio of thermonuclear reaction rates estimated with recommended rates given by Descouvemont {\it et al.}~\cite{Descouvemont:2004cw} [for $^7$Be($n,p$)$^7$Li] and Ando {\it et al.}~\cite{Ando:2005cz} [for $^1$H($n,\gamma$)$^2$H].  Because of a decrease in the $^7$Be($n,p$)$^7$Li rate at high energies, the ratio decreases at high temperatures.  At low temperatures ($T_9\lesssim 0.2$), $^7$Li is not destroyed, i.e., $P_{\rm des}(^7{\rm Li})\ll1$, although $^7$Be is transformed to $^7$Li via $^7$Be($n,p$)$^7$Li.  The amount of $^7$Li reduction is, therefore, small [Eq. (\ref{eq16})].  An efficient destruction of $^7$Li then prefers an operation of $^7$Be($n,p$)$^7$Li at higher temperature.  At high temperatures ($T_9\gtrsim 0.6$), on the other hand, the $^7$Be production in the SBBN is not yet completed.  Although $^7$Be nuclei are converted to $^7$Li, the same nuclei are produced via the reaction $^3$He($\alpha,\gamma$)$^7$Be later in lower temperatures until the reaction stops (Appendix~\ref{app2}).  In a white region at $0.2\lesssim T_9 \lesssim 0.6$, therefore, the reduction of $^7$Li is most efficient.


\begin{figure}
\begin{center}
\includegraphics[width=8.0cm,clip]{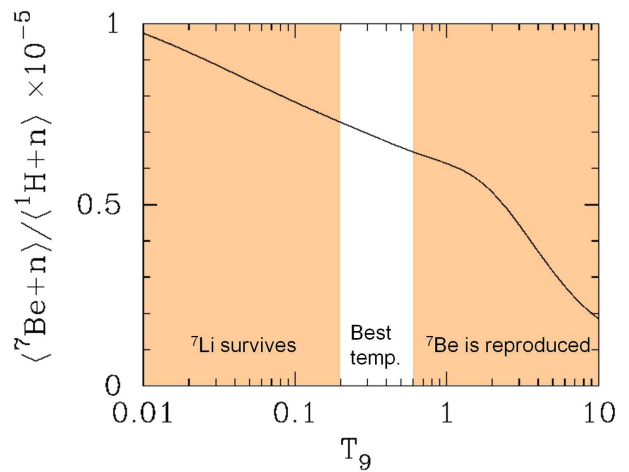}
\caption{Ratio between thermonuclear reaction rates of $^7$Be($n,p$)$^7$Li and $^1$H($n,\gamma$)$^2$H as a function of the temperature $T_9\equiv T/(10^9~{\rm K})$.  In a right shaded region at $T_9\gtrsim 0.6$, $^7$Be is reproduced through the reaction $^3$He($\alpha,\gamma$)$^7$Be after its destruction by neutron, while in a left shaded region at $T_9\lesssim 0.2$, the destruction of $^7$Li by the proton capture is inefficient.  The white region bounded by the shaded region, i.e., $0.2\lesssim T_9 \lesssim 0.6$, is the best temperature region in which extra neutrons efficiently reduce a final $^7$Li abundance.\label{fig1}}
\end{center}
\end{figure}


When energetic neutrons are injected, they experience an energy loss, especially the Coulomb scattering off the background electrons and positrons through interaction via their magnetic moments~\cite{Reno:1987qw,Kawasaki:2004qu,Jedamzik:2006xz}.  Nonthermal neutrons are then quickly thermalized.  Nevertheless, a small abundance of energetic neutrons can react with background H and $^7$Be before they could be thermalized.  At high neutron energies, the ratio of cross sections for $^7$Be($n,p$)$^7$Li and $^1$H($n,\gamma$)$^2$H is small.  Although the ratio of rates averaged over Maxwell-Boltzmann distribution is shown in Fig.~\ref{fig1}, the trend in reaction rate as a function of temperature roughly traces that in cross section as a function of energy.  Neutrons with higher energies thus relatively prefer the production of D over the destruction of $^7$Be.

In order to obtain a conservative lower limit on $\Delta {\rm D}/|\Delta ^7$Li$|$, we assume that nonthermal neutrons instantaneously thermalize, and cause a preferential reduction of $^7$Li.  Even if energetic hadrons induced by a hadronic energy injection were instantaneously thermalized, thermalized antinucleon can destroy background $^4$He nuclei through annihilation processes (Sec.~\ref{sec5}).

\section{Model (thermal neutron injection)}\label{sec3}

We assume that the TNI occurs at time $t_{\rm inj}$ instantaneously with a number density of injected neutron $\Delta n_{\rm inj}$. \footnote{Although the instantaneous injection is assumed in this study, injections of finite durations can be supposed.  The finite durations realize in spontaneous decays and annihilations of long-lived exotic particles or evaporations of exotic objects.}.  The abundance is measured as the number density relative to that of total baryons, i.e., $\Delta n_{\rm inj}/n_{\rm b}$.

\subsection{Method}\label{sec31}
We use the BBN code by Kawano~\cite{Kawano1992,Smith:1992yy} with the Sarkar's correction~\cite{Sarkar:1995dd} to $^4$He abundance.  Reaction rates relating to light nuclei of mass number $A \le  10$ are updated with the JINA REACLIB Database V1.0 \cite{2010ApJS..189..240C}.  We adopt the neutron lifetime of $878.5 \pm 0.7_{\rm stat} \pm 0.3_{\rm sys}$~s~\cite{Serebrov:2010sg}.

\subsection{Observational limits}\label{sec32}
We adopt an upper limit on the abundance ratio $^7$Li/H from a recent observation of MPSs, i.e., log($^7$Li/H)$=-12+(2.199\pm 0.086)$ derived with the 3D nonlocal thermal equilibrium model~\cite{Sbordone2010}.  Taking the two $\sigma$ (standard deviation) uncertainty, we assume the primordial abundance of $1.06\times 10^{-10} < {\rm ^7Li/H} < 2.35\times 10^{-10}$.  A consistency between a theoretical prediction and observations of MPSs requires a reduction of ($^7$Li+$^7$Be) during the BBN in amounts of at least $|\Delta ^7{\rm Li}|/{\rm H} \sim (5.24-2.35)\times 10^{-10}= 2.9\times 10^{-10}$.

The SBBN prediction of deuterium abundance is (D/H)$_{\rm SBBN}=2.56\times 10^{-5}$.  The final value of D/H$=({\rm D/H})_{\rm SBBN}+\Delta{\rm D/H}$ after the D production caused by the neutron injection should not deviate from primordial abundance inferred from observations of Lyman-$\alpha$ absorption system in the foreground of quasi-stellar objects (QSO).  Recent measurement of a damped Lyman $\alpha$ system QSO Sloan Digital Sky Survey (SDSS) J1419+0829 was performed most precisely of all QSO absorption systems ever found~\cite{Pettini:2012ph}.  We adopt the best measured abundance, log(D/H)=$-4.596\pm 0.009$ (best), and a mean value of ten QSO absorption line systems including J1419+0829, log(D/H)=$-4.58\pm 0.02$ (mean)~\cite{Pettini:2012ph}. \footnote{Recently Olive {\it et al.} \cite{Olive:2012xf} have considered possible effects of cosmic chemical evolution of D and $^7$Li, and suggested a solution to the $^7$Li problem by the $^7$Be destruction at the hadronic decay of a long-lived exotic particle followed by a depletion of D in the cosmic chemical evolution.}.

\section{Result}\label{sec4}
\begin{figure}
\begin{center}
\includegraphics[width=8.0cm,clip]{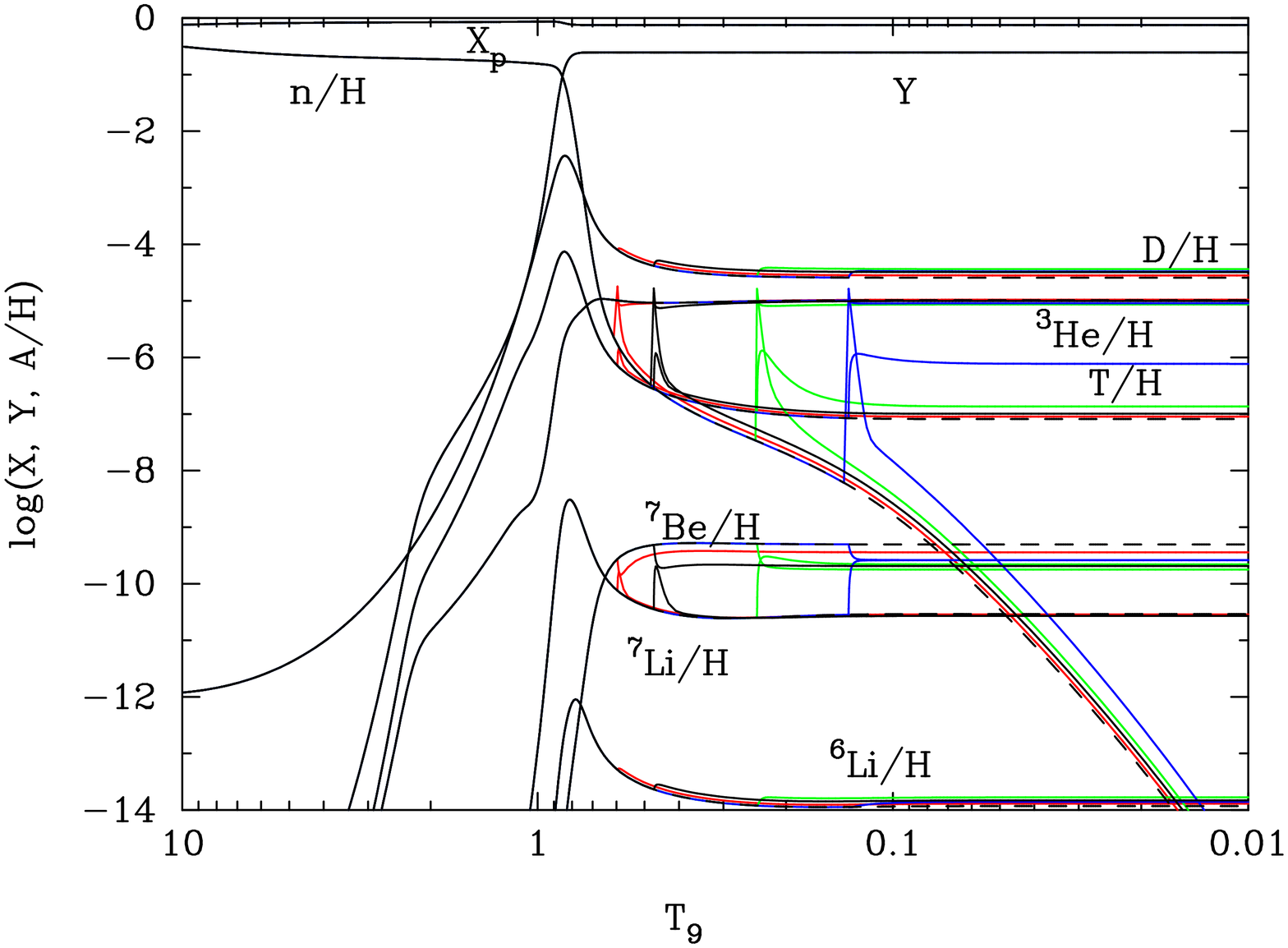}
\caption{Calculated abundances of H and $^4$He, i.e., $X$ and $Y$, respectively, (mass fractions), and other nuclides (number ratios relative to H) as a function of the temperature $T_9\equiv T/(10^9~{\rm K})$.  Solid lines are for cases of neutron injection by $\Delta n_{\rm inj}/n_{\rm b}=1.23\times 10^{-5}$ at $t_{\rm inj}=500$, $800$, $3000$, and $10^4$~s.  Dashed lines are for standard BBN model.  In all cases, final values of baryon-to-photon ratios are fixed to the WMAP value $\eta=6.2\times 10^{-10}$~\cite{Hinshaw:2012aka}.\label{fig2}}
\end{center}
\end{figure}

Figure \ref{fig2} shows calculated abundances of H and $^4$He, i.e., $X$ and $Y$, respectively, (mass fractions), and other nuclides (number ratios relative to H) as a function of the temperature $T_9$.  Solid lines correspond to cases of different injection times of $t_{\rm inj}=500$, $800$, $3000$, and $10^4$~s, for the same injected abundance $\Delta n_{\rm inj}/n_{\rm b}=1.23\times 10^{-5}$.  Dashed lines show fiducial abundances of the SBBN model.  In all cases, final values of baryon-to-photon ratios are the WMAP9 value $\eta=6.2\times 10^{-10}$ (model $\Lambda$CDM; WMAP data only)~\cite{Hinshaw:2012aka}.  In Appendix \ref{app1}, we describe important reactions through which nuclear abundances are affected.   It is seen that abundances of T, $^7$Li, and $^7$Be are changed much by the TNI, and that increases in abundances of T and $^7$Li depend significantly on $t_{\rm inj}$.  At a large value of $t_{\rm inj}$, the destruction reaction of T, i.e., $^3$H($d$, $n$)$^4$He, is ineffective because of a low temperature.  The final T abundance is then large.  T nuclei produced during the BBN epoch decay to $^3$He with the half life of $12.32 \pm 0.02$ yr \cite{2010NuPhA.848....1P}.  The final $^3$He abundance is, therefore, given by a sum of abundances of T and $^3$He at BBN.  Since the $^3$He abundance is much larger than the T abundance in the BBN epoch, increases of T abundance change the final $^3$He abundance by only negligible amounts.


\begin{figure}
\begin{center}
\includegraphics[width=8.0cm,clip]{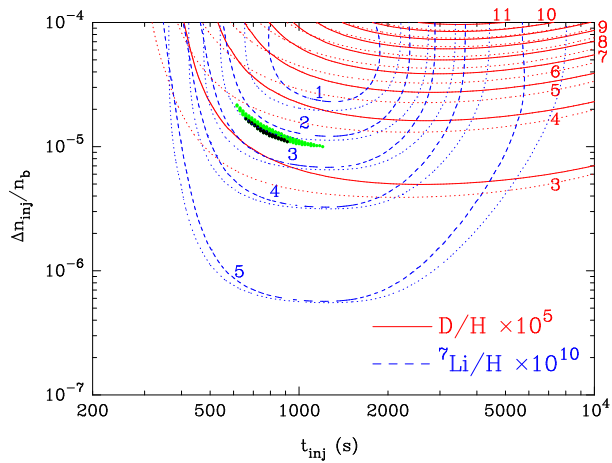}
\caption{Contours for final abundances of D (solid lines) and $^7$Li (dashed lines) in the ($t_{\rm inj}$, $\Delta n_{\rm inj}/n_{\rm b}$) plane.  Analytical estimates with Eqs. (\ref{eqb4}), (\ref{eqb7}), (\ref{eqb9}), and (\ref{eqb11}) are also shown by thin dotted lines.  Points at $(t_{\rm inj}$, $\Delta n_{\rm inj}/n_{\rm b})\sim (800$~s, $10^{-5}$) indicate parameter sets which reproduce the observed $^7$Li abundance in the 2 $\sigma$ range~\cite{Sbordone2010} while keeping D abundance in the 12 $\sigma$ range of the best value [dark (black) points], and the 5 $\sigma$ range of the mean value [light (green) points], respectively~\cite{Pettini:2012ph}.
\label{fig3}}
\end{center}
\end{figure}

Figure \ref{fig3} shows contours for final abundances of D (solid lines) and $^7$Li (dashed lines) in the ($t_{\rm inj}$, $\Delta n_{\rm inj}/n_{\rm b}$) plane.  In a narrow region indicated at $(t_{\rm inj}$, $\Delta n_{\rm inj}/n_{\rm b})\sim (800$~s, $10^{-5}$) by points, the primordial abundance inferred from observations of $^7$Li~\cite{Sbordone2010} is reproduced within the two $\sigma$ uncertainty keeping the D abundances close to the observed value~\cite{Pettini:2012ph}.  This region is, therefore, the most preferred region.  Dark (black) points correspond to calculated D abundances in the 12 $\sigma$ range of the best observed value, while light (green) points correspond to those in the 5 $\sigma$ range of the mean value.  It is found that D abundances in the 11 $\sigma$ range of the best value and the 4 $\sigma$ range of the mean value are never accompanied with Li abundances in the observational 2 $\sigma$ range in this model.  The recent precise determination of D abundance in the QSO absorption line systems thus completely excludes the solution to the Li problem in this model.  This calculation itself should be similar to a recent calculation for neutron injection which concluded that this model can provide a solution to the Li problem \cite{Vasquez:2012dz}.  Our different conclusion results from the use of the new observational constraints on primordial D abundance.  Effects on abundances of $n$, D, $^3$H, $^3$He, $^6$Li, $^7$Li, and $^7$Be are different in different parameter cases.  Reasons for that are described in Appendixes \ref{app1} and \ref{app2}.


\begin{figure}
\begin{center}
\includegraphics[width=8.0cm,clip]{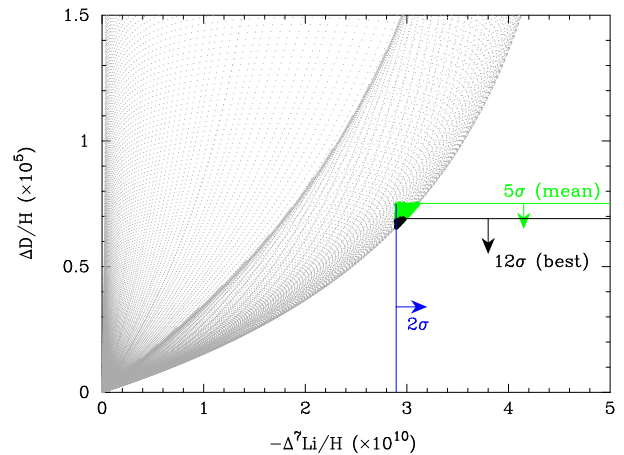}
\caption{Differences between abundances in BBN with neutron injections and those in standard BBN for D (vertical axis) and $^7$Li (horizontal).  The lines with arrows indicate the parameter regions which satisfy the 12$\sigma$ constraint (best) and the 5 $\sigma$ constraint (mean) of D abundance and the 2$\sigma$ constraint of $^7$Li abundance.  The dark (black) and light (green) points inside the region bounded by two lines correspond to parameter sets satisfying those constraint regions shown in Fig. \ref{fig3}.
\label{fig4}}
\end{center}
\end{figure}

Figure \ref{fig4} shows a region on the parameter plane of ($-\Delta ^7$Li/H, $\Delta$D/H) which can be occupied in this model.   The lines with arrows indicate the regions which satisfy observational constraints on abundances of D (12 $\sigma$ for the best value, and 5 $\sigma$ for the mean value)~\cite{Pettini:2012ph} and $^7$Li (2 $\sigma$) \cite{Sbordone2010}.  A lower limit on $\Delta {\rm D}/{\rm H}$ as a function of $-\Delta ^7$Li/H can be read from this figure.  The points at ($-\Delta ^7$Li/H, $\Delta$D/H) $\sim (3\times 10^{-10}$, $0.7\times 10^{-5}$) satisfy the constraints.  Abundances in this parameter region give close agreement with those found in a recent detailed study on effects of hadronic decay~\cite{Cyburt:2010vz} as their most favorable results of abundances.

In our preferred parameter region, the abundance of D is related to that of $^7$Li for the adopted nuclear reaction rates (Sec.~\ref{sec31}) and baryon-to-photon ratio~\cite{Hinshaw:2012aka} as described by
\begin{equation}
 \frac{\rm D}{\rm H}~>\left[2.56+0.227\left(5.24-\frac{^7{\rm Li}}{\rm H}\times 10^{10}\right)\right]\times 10^{-5}.
\label{eq17}
\end{equation}
This constraint is free of many uncertainties related to nuclear and electromagnetic reactions for nonthermal particles produced by the neutron injection.  Although primary antinucleons, and secondary and higher order neutrons always increase the ratio $\Delta$D/$|\Delta ^7$Li$|$ (see Secs.~\ref{sec22} and \ref{sec5}), their effects depend [Eqs. (\ref{eq15}) and (\ref{eq18}))] on information of relative injected amounts of $n$, $p$, $\bar{n}$ and $\bar{p}$, and their injected energy spectra.  The information itself depends on the decay property of the long-lived exotic particle such as its mass and decay modes.  Equation (\ref{eq17}) then corresponds to the most conservative model independent lower limit on D/H as a function of $^7$Li/H.


\subsection{$t_{\rm inj}$ dependence}\label{sec41}

In the case of earliest neutron injection at $t_{\rm inj}=500$~s (Fig.~\ref{fig2}), effects of additional neutrons are removed by efficient nuclear reactions.  Especially, although the $^7$Be abundance reduces right after the neutron injection, the reaction $^3$He($\alpha,\gamma$)$^7$Be enhances $^7$Be again.

In the best case of injection, i.e., $t_{\rm inj}=800$~s, the $^7$Be abundance decreases and the D abundance increases a little less efficiently.

In the case of later injection, i.e., $t_{\rm inj}=3000$~s, the D abundance increases via $^1$H($n,\gamma$)$^2$H, and is not affected by already inefficient D destruction reactions.  The resulting D abundance is thus larger than in the best case.  The $^7$Be conversion to $^7$Li by neutron capture efficiently proceeds.  However, the reaction $^7$Li($p,\alpha$)$^4$He is no longer operative.  The resulting decrease in the mass-number-seven ($^7$Li+$^7$Be) abundance is, therefore, very small.

In the case of the latest injection, i.e., $t_{\rm inj}=10^4$~s, some portion of injected neutrons decay with the lifetime $\tau_n=878.5$~s~\cite{Serebrov:2010sg} before they could trigger the D production via $^1$H($n,\gamma$)$^2$H.  This leads to a suppressed D production.  The reduction of ($^7$Li+$^7$Be) is not operative as in the previous case.

If the neutron injection has a duration, deviations in final abundances would be approximately given by weighted average over time of deviations obtained in this instantaneous injection model.

\subsection{$\Delta n_{\rm inj}/n_{\rm b}$ dependence}\label{sec42}
When amounts of neutron injection are small, i.e., $\Delta n_{\rm inj}/n_{\rm b}\lesssim 10^6$, both of the $^7$Be destruction and the D production are efficient.  If the injection is strong, i.e, $\Delta n_{\rm inj}/n_{\rm b}\gtrsim 10^5$, however, the efficiency of $^7$Be destruction plateaus since it gets difficult for neutrons to find $^7$Be nuclei with an already small abundance~[cf. Eq. (\ref{eqb7})].  The efficiency in the D production, on the other hand, is not suppressed since the target of neutrons at the reaction $^1$H($n,\gamma$)$^2$H is proton whose abundance is very large, and dose not change significantly in this model for parameter values of $\Delta n_{\rm inj}/n_{\rm b}$ and $t_{\rm inj}$ considered here.

\section{Antinucleon+$^4$H\lowercase{e} annihilation}\label{sec5}

The antinucleon ($\bar{N}$)+$^4$He annihilation (as considered in Ref.~\cite{1982NCimR...5j...1C}) is an important process which always operates when $\bar{N}$'s are produced.  The annihilation of (thermalized) $\bar{N}$ and $^4$He affects the elemental abundances even when productions of secondary particles via $^4$He spallations by energetic hadrons can be neglected.  The annihilations produce light mesons, $n$, $p$, $d$, $t$, and $^3$He.

Generated neutrons of abundance $\Delta n_{\rm inj}$ are almost completely captured by protons, and produce deuterons if the time of neutron injection is $t_{\rm inj}\lesssim 10^3$~s.  The final abundance of D produced through the $\bar{N}$+$^4$He annihilation is then given by
\begin{equation}
\frac{\Delta {\rm D}_1^{\rm ann}}{\rm H}=\sum_{\bar{N}=\bar{n},\bar{p}} \frac{\bar{N}_1}{\rm H}\left(\frac{n_\alpha\sigma_\alpha}{n_{\rm H}\sigma_p+n_\alpha\sigma_\alpha}\right)_{\bar{N}}P_d(\bar{N})P_{\rm sur}({\rm D}),
\label{eq18}
\end{equation}
where
$\bar{N}_1/$H is the number densities of primary $\bar{N}$ injected simultaneously at the injection of neutrons [cf. Eq. (\ref{eq9})] relative to that of background hydrogen.  The ratio in the parenthesis with subscript $\bar{N}$ is the value for annihilation of species $\bar{N}$ [cf. Eq. (\ref{eq12})].  $P_d(\bar{N})$ is the fraction of annihilation into final states including a deuteron to that for total annihilation.

Equation (\ref{eq18}) is transformed to an equation:
\begin{eqnarray}
\frac{\Delta {\rm D}_1^{\rm ann}}{\rm H}&=&\frac{\Delta n_{\rm inj}}{n_{\rm H}}~\frac{1}{1+\sum_{\bar{N}}f_{\rm add}(\bar{N})}\nonumber\\
&&\times N_{\rm eff}\left[\left(\frac{n_\alpha\sigma_\alpha}{n_{\rm H}\sigma_p+n_\alpha\sigma_\alpha}\right)_{\bar{N}}~P_d(\bar{N})\right]P_{\rm sur}({\rm D}), \nonumber\\
\label{eq19}
\end{eqnarray}
where
$\Delta n_{\rm inj}/n_{\rm H}$ is the number abundance of generated neutron relative to that of $^1$H.
$N_{\rm eff}\equiv (\bar{n}_1+\bar{p}_1)/n_1$ is the effective number of primary antinucleons per primary neutron, and
$f_{\rm add}(\bar{N})$ is the number ratio between the neutron produced secondarily by the annihilation of $\bar{N}$ plus $^4$He, and the primary neutron.  The square bracket in the second line is the quantity averaged over $\bar{N}=\bar{n}$ and $\bar{p}$ with weights of $\bar{N}_1$.
The equation, i.e., $\Delta n_{\rm inj}=n_1 (1+\sum_{\bar{N}}f_{\rm add}(\bar{N}))$, is satisfied.

We try an example estimation.  We assume that abundances of nonthermal primary antinucleons are twice as large as those of primary neutron.  This leads to $N_{\rm eff}=2$.  We assume $P_d(\bar{N})=0.1$, $P_n(\bar{N})<0.4$ \cite{1988NCimA.100..323B} for both $\bar{n}$ and $\bar{p}$, $\sigma_\alpha/\sigma_p=4^{2/3}$~\cite{Levitan:1988au}, and $n_\alpha/n_{\rm H}=0.082$, as done in deriving Eq. (\ref{eq13}).  The following equation is then derived:
\begin{eqnarray}
\sum_{\bar{N}} f_{\rm add}(\bar{N})&=&N_{\rm eff}\left(\frac{n_\alpha\sigma_\alpha}{n_{\rm H}\sigma_p+n_\alpha\sigma_\alpha}\right)_{\bar{N}}~P_n(\bar{N})<0.14. \nonumber\\
\label{eq20}
\end{eqnarray}
Using Eqs.~(\ref{eq19}) and (\ref{eq20}), we obtain
\begin{eqnarray}
\frac{\Delta {\rm D}_1^{\rm ann}}{\rm H}> 0.030\frac{\Delta n_{\rm inj}}{X n_{\rm b}}P_{\rm sur}({\rm D}).
\label{eq21}
\end{eqnarray}

This component should add to the production of D in the present model in which only effects of neutron were taken into account.  The total change of D abundance is, therefore, given by $\Delta {\rm D/H}=[\Delta n_{\rm inj}/(X n_{\rm b})]P_{\rm sur}({\rm D})+\Delta {\rm D_1^{\rm ann}}/{\rm H}>1.03 [\Delta n_{\rm inj}/(X n_{\rm b})]P_{\rm sur}({\rm D})$.  In this case, the abundance of D in the preferred parameter region (Sec.~\ref{sec4}) is
\begin{equation}
 \frac{\rm D}{\rm H}~>\left[2.56+0.234\left(5.24-\frac{^7{\rm Li}}{\rm H}\times 10^{10}\right)\right]\times 10^{-5}.
\label{eq22}
\end{equation}

\begin{figure}
\begin{center}
\includegraphics[width=8.0cm,clip]{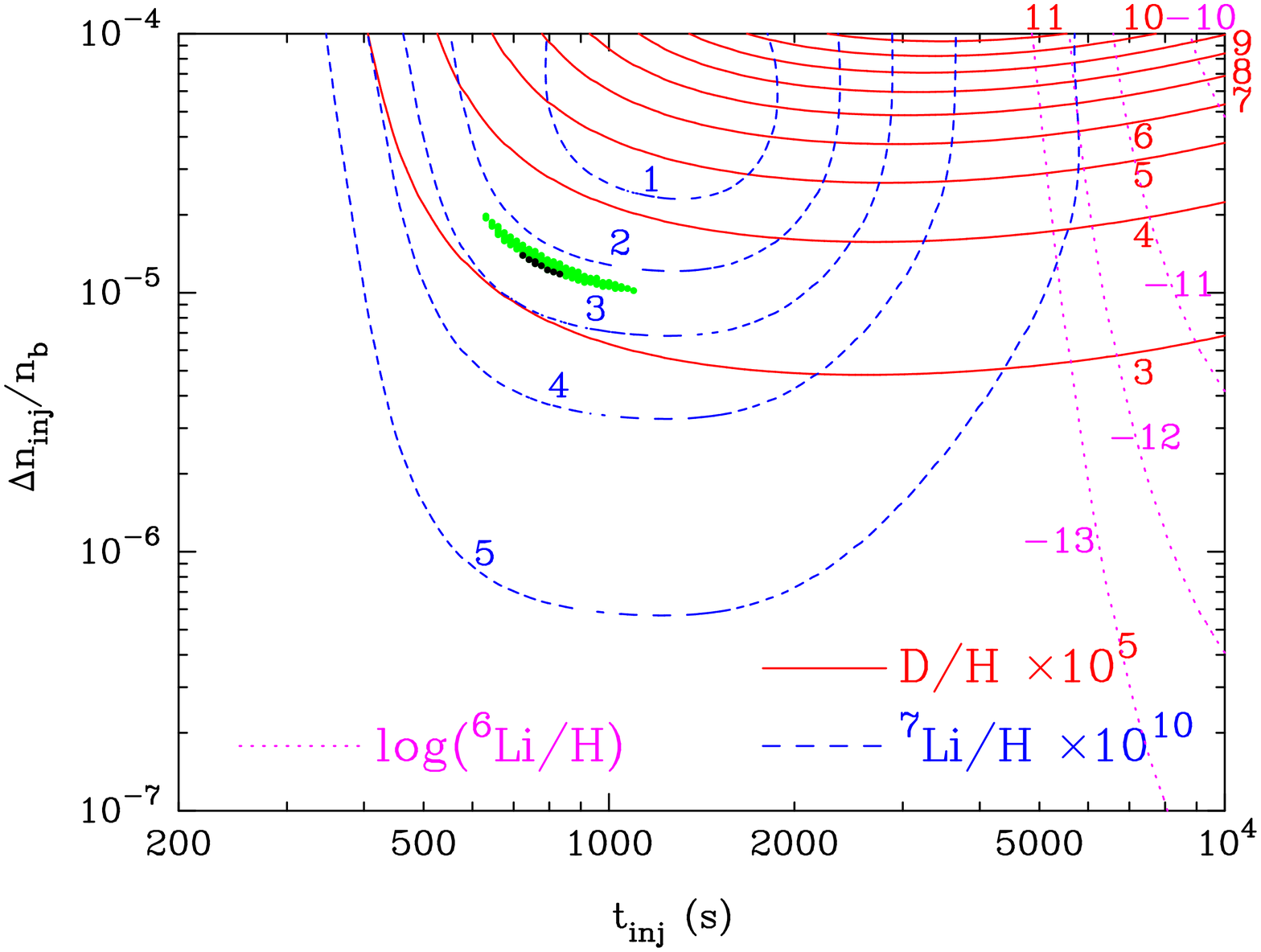}
\caption{Same as in Fig. \ref{fig3} for the case that effects of the antinucleon+$^4$He annihilation are taken into account as described in Sec.~\ref{sec5}.  Parameter regions reproducing observed abundances of D~\cite{Pettini:2012ph} and $^7$Li~\cite{Sbordone2010} are narrower than those in Fig. \ref{fig3}.  Dotted lines are contours for final abundance of $^6$Li for the case that $^6$Li production through secondary reactions triggered by antinucleon is taken into account (see Sec. \ref{sec6}).
\label{fig5}}
\end{center}
\end{figure}

Figure \ref{fig5} shows contours for final abundances of D (solid lines) and $^7$Li (dashed lines) on the ($t_{\rm inj}$, $\Delta n_{\rm inj}/n_{\rm b}$) plane in the case that the additional D production from the annihilation is taken into account by the lower limit, i.e., Eq. (\ref{eq21}).  We find that this small fraction of additional D production narrows the best parameter region in Figs.~\ref{fig3} and \ref{fig4} without moving contours in Fig. \ref{fig3} significantly.


\section{$^6$L\lowercase{i} production from primary antinucleons}\label{sec6}
The $\bar{N}+^4$He annihilation produces nonthermal $^3$H and $^3$He.  The nuclides with mass number three can react with background $^4$He, and produce $^6$Li.  The decay of $^3$H can be neglected since its half life, i.e., $12.3$~y~\cite{2010NuPhA.848....1P}, is much longer than time scales of related processes [e.g., inverse of Eq. (\ref{eq4})] in the relevant temperature range.  The abundance from this $\bar{N}$-induced $^6$Li production is then estimated as
\begin{eqnarray}
\frac{\Delta ^6{\rm Li}}{\rm H}&\hspace{-5pt}=&\hspace{-5pt}
\left[\sum_{\bar{N}=\bar{n},\bar{p}} \frac{\bar{N}_1}{\rm H}\left(\frac{n_\alpha\sigma_\alpha}{n_{\rm H}\sigma_p+n_\alpha\sigma_\alpha}\right)_{\bar{N}}~\sum_{A=t,^3{\rm He}}~P_{A}(\bar{N}) \right.\nonumber\\
&\hspace{-5pt}&\hspace{-5pt}\left. \times \int_{E_{{\rm th},A}}^\infty~f_A(\bar{N},E_A)~\frac{n_\alpha\sigma_{A(\alpha,N)^6{\rm Li}}~v_A}{\Gamma_{\rm tot}(E_A)} dE_A\right] \nonumber\\
&\hspace{-5pt}&\hspace{-5pt}\times P_{\rm sur}(^6{\rm Li}),
\label{eq23}
\end{eqnarray}
where
$P_A(\bar{N})$ is the ratio of the cross section for annihilation into final states including a nuclide $A$ to that for total annihilation.
$\sigma_{A(\alpha,N)^6{\rm Li}}$ and $E_{{\rm th},A}$ are the cross section and the threshold energy for the reaction $A$($\alpha,N$)$^6$Li,
$E_A$ and $v_A$ are the kinetic energy and the velocity of $A$, and
$f_A(\bar{N},E_A)$ is the distribution function of secondary $A$ produced at the annihilation of $\bar{N}$+$^4$He as a function of $E_A$.
$\Gamma_{\rm tot}=\Gamma_{\rm inel}+\Gamma_{\rm loss}+...$ is the total reaction rate of nuclide $A$ as a function of $E_A$, and
$P_{\rm sur}(^6{\rm Li})$ is the survival fraction of $^6$Li, which is produced via the secondary reaction $A$($\alpha,N$)$^6$Li, during its propagation.

We try an example estimation for this component of nonthermal $^6$Li production.  We assume $N_{\rm eff}=2$, $P_n(\bar{N})<0.4$, $P_t(\bar{N})=0.44$ and $P_{^3{\rm He}}=0.21$~\cite{1988NCimA.100..323B} for both $\bar{n}$ and $\bar{p}$, $\sigma_\alpha/\sigma_p=4^{2/3}$~\cite{Levitan:1988au}, and $n_\alpha/n_{\rm H}=0.082$, as done in Sec. \ref{sec5}.  The energy spectra of mass-three-nuclides, i.e., $f_A(\bar{N},E_A)$ were assumed to be given by an result of experiment measuring the spectrum for $^3$He at $\bar{p}+^4$He annihilation~\cite{1988NCimA.100..323B}.  In the experiment, no dependence of the spectrum on the initial $\bar{p}$ energy has been observed, and the nuclide $^3$He in the final state can be identified without being confused with other hadronic species.

The reaction cross sections $\sigma_{A(\alpha,N)^6{\rm Li}}$ are taken from Ref. \cite{Cyburt:2002uv}.  The total rate $\Gamma_{\rm tot}$ is assumed to be the Coulomb loss rate since the Coulomb loss dominates as long as the energy is not too high.  At temperature $T\leq m_e$, the Coulomb loss rate of relativistic charged particles is given by Eq. (\ref{eq3}).  The rate of non-relativistic charged particles is given~\cite{Reno:1987qw,Kawasaki:2004qu} by
\begin{eqnarray}
 \frac{dE}{dt}&\hspace{-5pt}=&\hspace{-5pt}-\frac{4\pi Z^2 \alpha^2}{m_e v}\Lambda~n_e\left[\frac{4}{\pi^{1/2}} \int_0^{v/\sqrt{\mathstrut 2T/m_e}} x^2 \exp(-x^2)~dx \right.\nonumber\\
&\hspace{-5pt}&\hspace{-5pt}\left.~~~~~~~~~~~~~~+ \frac{2^{1/2}}{3\pi^{1/2}} \sqrt{\mathstrut \frac{m_e}{T}} v^3 \exp\left(-\frac{m_e v^2}{2T}\right)\right].
\label{eq24}
\end{eqnarray}
The $^6$Li survival fraction $P_{\rm sur}(^6{\rm Li})$ is calculated in our BBN code.

In Fig. \ref{fig5}, dotted lines correspond to abundance ratios, i.e., $^6$Li/H$=(^6$Li/H$)_{\rm SBBN}+(\Delta ^6$Li/H$)_{t_{\rm inj}}P_{\rm sur}(^6{\rm Li})=10^{-13}$, $10^{-12}$, $10^{-11}$, and $10^{-10}$ (from bottom to top).  In high temperature environments, $^6$Li nuclei produced in the reaction $A$($\alpha,N$)$^6$Li are effectively destroyed via proton burning, i.e., $P_{\rm sur}(^6{\rm Li})\ll 1$.  A significant production of $^6$Li then occurs at relatively low temperature when the $^6$Li destruction is ineffective and the energy loss rate of secondary nuclides $^3$H and $^3$He through Coulomb scattering off background $e^\pm$ is diminished because of the reduced abundances of $e^\pm$ through their pair annihilation.

\section{Conclusions}\label{sec7}
The injections of energetic hadrons could have occurred in the early universe by hypothetical events of decays or annihilations of long-lived exotic particles, or evaporations of exotic objects.  The injections cause scattering of thermal nuclei by energetic hadrons, and showers of nonthermal nucleons, antinucleons, and nuclei can develop.  Neutrons generated at the exotic events can react with $^7$Be and reduce final abundances of $^7$Li (which are mainly produced via the electron capture of $^7$Be).  It has been suggested that the $^7$Be reduction can be a solution to a discrepancy between theoretical $^7$Li abundances of the SBBN model and that inferred from observations of Galactic metal-poor stars.  The theoretical abundance is about a factor of three larger than the observational one.

Based on an analysis of related physical processes, we prove that the assumption of instantaneous thermalization of injected neutron provides the way to derive a conservative limit on the relation between abundances of D and $^7$Li in the hadronic energy injection model, which is independent of uncertainties in generations and reactions of nonthermal hadrons originating from the injections (Sec.~\ref{sec2}).  Furthermore, two important points are stressed:  1) An uncertainty in cross sections of inelastic $n+p$ scattering [Eqs. (\ref{eq6}), (\ref{eq7}), and (\ref{eq8})] affects the total number of neutrons generated from the primary neutron injection, which is critical for resulting abundances of D and $^7$Li.  2) One must include effects of annihilations of antinucleons with $^4$He on a primordial D abundance even if antinucleons generated with neutron were instantaneously thermalized.

We then consider a simple model in which extra thermal neutrons are injected in a late epoch of the BBN.  We estimate the probability that primordial abundances of $^7$Li in this model can be consistent with observed abundances.  Relations between primordial abundances of D and $^7$Li are obtained in a manner to conserve the probability securely.

We perform a BBN calculation, and find a very small parameter region of the neutron injection time ($t_{\rm inj}$) and the number density ($\Delta n_{\rm inj}$) of injected neutron in which $^7$Li abundances are within the 2 $\sigma$ uncertainty range determined from observation and changes in D abundance are minimum.  In the preferred parameter region, the injection time is $t_{\rm inj}\sim 800$~s, and its number density is $10^{-5}$ times as large as that of total baryonic matter.  A typical pattern of nucleosynthesis in the parameter region is analyzed (Appendix~\ref{app1}).  Situations of D production and $^7$Li reduction are observed especially (Appendix~\ref{app2}).

We derive a model-independent result [Eq.~(\ref{eq17})] that a reduction of $^7$Li abundance from the SBBN value down to the observational two $\sigma$ upper limit is necessarily accompanied by an undesirable increase of D abundance up to at least the 12 $\sigma$ upper limit (best observed value) and the 5 $\sigma$ upper limit (mean observed value).  When effects of antinucleons+$^4$He annihilations are considered utilizing a possible example case, the preferred parameter regions become narrower in the present model.  BBN models involving any injections of extra neutron are, therefore, not likely to accommodate alone a reduction of primordial $^7$Li abundance to the observed level.

\appendix

\section{Important reactions}\label{app1}

We analyzed nucleosynthesis with a BBN code, and found important reactions operating in the case of extra neutron injection of $\Delta n_{\rm inj}/n_{\rm b}=10^{-5}$ at $t_{\rm inj}=1100$~s corresponding to $T_9=0.4$ (see Fig. \ref{fig2}).  We list rates of dominant reactions for productions ($\Gamma_{\rm pro}$) and destructions ($\Gamma_{\rm des}$) of respective nuclides.

Nuclear reactions do not operate effectively if the rates are smaller than the cosmic expansion rate given by
\begin{eqnarray}
H&=&\frac{2}{3\sqrt[]{\mathstrut 5}} \frac{\pi^{3/2}}{m_{\rm Pl}}
 g_\ast^{1/2} T^2=4.5(-4)~{\rm s}^{-1}~g_{\ast,3.36}^{1/2}~T_{9,0.4}^2.~~~~~~~
\label{eqa1}
\end{eqnarray}
where
$m_{\rm Pl}$ is the Planck's mass, and
$g_\ast$ is the total number of effective massless degrees of freedom \cite{kolb1990}.

In what follows, $\langle 1(2,3)4 \rangle$ denotes the rate, i.e., $\langle \sigma v \rangle$ for a reaction 1(2,3)4 with the cross section $\sigma$, and the relative velocity $v$.  Rates are measured in the unit of cm$^3$~s$^{-1}$~mol$^{-1}$.

\subsection{$n$}
An instantaneous production of extra neutron has been assumed:
\begin{eqnarray}
\Gamma_{\rm des}&=&n_{\rm H} \langle ^1{\rm H}(n,\gamma)^2{\rm H} \rangle \nonumber\\
&=& 3.1(-2)~{\rm s}^{-1}~T_{9,0.4}^3~\eta_{,6.2(-10)}~X_{,0.75}\langle \sigma v \rangle_{,3.1(4)}.\nonumber\\
\label{eqa2}
\end{eqnarray}

\subsection{D}
\begin{eqnarray}
\Gamma_{\rm pro}&=&\frac{n_{\rm H} n_n}{n_{\rm D}} \langle ^1{\rm H}(n,\gamma)^2{\rm H} \rangle \nonumber\\
&=& 9.1(-3)~{\rm s}^{-1}~T_{9,0.4}^3~\eta_{,6.2(-10)}~X_{,0.75}\nonumber\\
&& \times n/{\rm H}_{,1(-5)}~\left({\rm D/H}_{,3.4(-5)}\right)^{-1}~ \langle \sigma v \rangle_{,3.1(4)}.\nonumber\\
\label{eqa3}
\end{eqnarray}
\begin{eqnarray}
\Gamma_{\rm des}&=&n_{\rm D} \langle ^2{\rm H}(d,N)^3A \rangle \nonumber\\
&=& 2.6(-4)~{\rm s}^{-1}~T_{9,0.4}^3~\eta_{,6.2(-10)}~X_{,0.75}~{\rm D/H}_{,3.4(-5)} \nonumber\\
&&\times \langle \sigma v \rangle_{,7.5(6)},
\label{eqa4}
\end{eqnarray}
where
$N=n$ or $p$, and
$^3A=^3$He or $t$.
Rates for final states of $n+^3$He and $p+^3$H are $4.0\times 10^6$ and $3.5\times 10^6$ cm$^3$s$^{-1}$mol$^{-1}$, respectively.

\subsection{$^3$H}
\begin{eqnarray}
\Gamma_{\rm pro}&=&\frac{n_{\rm D}^2}{n_{^3{\rm H}}} \langle  ^2{\rm H}(d,p)^3{\rm H} \rangle \nonumber\\
&=& 2.1(-2)~{\rm s}^{-1}~T_{9,0.4}^3~\eta_{,6.2(-10)}~X_{,0.75}\nonumber\\
&& \times \left({\rm D/H}_{,3.4(-5)}\right)^2~\left(^3{\rm H/H}_{,1.9(-7)}\right)^{-1}~ \langle \sigma v \rangle_{,3.5(6)}.\nonumber\\
\label{eqa5}
\end{eqnarray}
\begin{eqnarray}
\Gamma_{\rm des}&=&n_{\rm D} \langle ^3{\rm H}(d,n)^4{\rm He} \rangle \nonumber\\
&=& 1.5(-2)~{\rm s}^{-1}~T_{9,0.4}^3~\eta_{,6.2(-10)}~X_{,0.75}~{\rm D/H}_{,3.4(-5)} \nonumber\\
&&\times \langle \sigma v \rangle_{,4.3(8)}.
\label{eqa6}
\end{eqnarray}
The rate for $^3$He($d,p$)$^4$He is about 30 times smaller than that for $^3$H($d,n$)$^4$He.

\subsection{$^3$He}
\begin{eqnarray}
\Gamma_{\rm pro}&=&\frac{n_{\rm D}^2}{n_{^3{\rm He}}} \langle ^2{\rm H}(d,n)^3{\rm He} \rangle \nonumber\\
&=& 5.0(-4)~{\rm s}^{-1}~T_{9,0.4}^3~\eta_{,6.2(-10)}~X_{,0.75}\nonumber\\
&& \times \left({\rm D/H}_{,3.4(-5)}\right)^2~\left(^3{\rm He/H}_{,9.3(-6)}\right)^{-1}~ \langle \sigma v \rangle_{,4.0(6)}.\nonumber\\
\label{eqa7}
\end{eqnarray}
\begin{eqnarray}
\Gamma_{\rm des}&=&n_n \langle ^3{\rm He}(n,p)^3{\rm H} \rangle \nonumber \\
&=& 5.2(-3)~{\rm s}^{-1}~T_{9,0.4}^3~\eta_{,6.2(-10)}~X_{,0.75}~n/{\rm H}_{,1(-5)} \nonumber\\
&&\times \langle \sigma v \rangle_{,5.2(8)}.
\label{eqa8}
\end{eqnarray}

\subsection{$^6$Li}
\begin{eqnarray}
\Gamma_{\rm pro}&=&\frac{n_{^4{\rm He}} n_{\rm D}}{n_{^6{\rm Li}}} \langle ^4{\rm He}(d,\gamma)^6{\rm Li} \rangle \nonumber\\
&=& 6.8(-1)~{\rm s}^{-1}~T_{9,0.4}^3~\eta_{,6.2(-10)}~Y_{,0.25}\nonumber\\
&& \times {\rm D/H}_{,3.4(-5)}~\left(^6{\rm Li/H}_{,1.7(-14)}\right)^{-1}~ \langle \sigma v \rangle_{,4.1(-3)}.\nonumber\\
\label{eqa9}
\end{eqnarray}
\begin{eqnarray}
\Gamma_{\rm des}&=&n_{\rm H} \langle ^6{\rm Li}(p,\alpha)^3{\rm He} \rangle \nonumber\\
&=& 6.5(-1)~{\rm s}^{-1}~T_{9,0.4}^3~\eta_{,6.2(-10)}~X_{,0.75}~ \langle \sigma v \rangle_{,6.5(5)}.\nonumber\\
\label{eqa10}
\end{eqnarray}
The destruction and production of $^6$Li still operate efficiently in this low temperature environment.  The abundance of $^6$Li is, therefore, the steady state abundance determined from $\Gamma_{\rm pro}=\Gamma_{\rm des}$.

\subsection{$^7$Li}
\begin{eqnarray}
\Gamma_{\rm pro}&=&\frac{n_{^4{\rm He}} n_{^3{\rm H}}}{n_{^7{\rm Li}}} \langle ^4{\rm He}(t,\gamma)^7{\rm Li} \rangle \nonumber\\
&=& 1.2(-2)~{\rm s}^{-1}~T_{9,0.4}^3~\eta_{,6.2(-10)}~Y_{,0.25}\nonumber\\
&& \times ^3{\rm H/H}_{,1.9(-7)}~\left(^7{\rm Li/H}_{,2.8(-11)}\right)^{-1}~ \langle \sigma v \rangle_{,2.2(1)}.\nonumber\\
\label{eqa11}
\end{eqnarray}
\begin{eqnarray}
\Gamma_{\rm des}&=&n_{\rm H} \langle ^7{\rm Li}(p,\alpha)^4{\rm He} \rangle \nonumber\\
&=& 1.8(-2)~{\rm s}^{-1}~T_{9,0.4}^3~\eta_{,6.2(-10)}~X_{,0.75}~ \langle \sigma v \rangle_{,1.8(4)}.\nonumber\\
\label{eqa12}
\end{eqnarray}
$^7$Li also experiences destruction and production efficiently. The $^7$Li abundance is the steady state abundance.

\subsection{$^7$Be}
\begin{eqnarray}
\Gamma_{\rm pro}&=&\frac{n_{^4{\rm He}} n_{^3{\rm He}}}{n_{^7{\rm Be}}} \langle ^4{\rm He}(^3{\rm He},\gamma)^7{\rm Be} \rangle \nonumber\\
&=& 3.9(-4)~{\rm s}^{-1}~T_{9,0.4}^3~\eta_{,6.2(-10)}~Y_{,0.25}\nonumber\\
&& \times ^3{\rm He/H}_{,9.3(-6)}~\left(^7{\rm Be/H}_{,5.2(-10)}\right)^{-1}~ \langle \sigma v \rangle_{,2.6(-1)},\nonumber\\
\label{eqa13}
\end{eqnarray}
\begin{eqnarray}
\Gamma_{\rm des}&=&n_n \langle ^7{\rm Be}(n,p)^7{\rm Li} \rangle \nonumber\\
&=&2.1(-2)~{\rm s}^{-1}~T_{9,0.4}^3~\eta_{,6.2(-10)}~X_{,0.75}~n/{\rm H}_{,1(-5)} \nonumber\\
&&\times \langle \sigma v \rangle_{,2.1(9)},
\label{eqa14}
\end{eqnarray}
$^7$Be is only transformed into $^7$Li, and the $^7$Li abundance instantaneously relaxes to the steady state abundance.

\section{Analytical estimates}\label{app2}
\subsection{D production}
The evolution of extra neutron abundance, i.e., $\Delta n$, is described simply by
\begin{equation}
\frac{d \Delta n}{dt}=-3H \Delta n -n_{\rm H} \Delta n \langle {\rm H}+n \rangle -\frac{\Delta n}{\tau_n},
\label{eqb1}
\end{equation}
where
the first, second, and third terms of the RHS correspond to the dilution by cosmic expansion, the reduction by the radiative proton capture, and the reduction by $\beta$-decay, respectively.  $\langle i+j \rangle$ denotes the reaction rate, i.e., $\langle \sigma v \rangle$, for a reaction of species $i$ and $j$.  For a simplistic understanding, we assume that the destruction by the radiative capture reduces the neutron abundance instantaneously compared to the time scale of Hubble expansion, and that the neutron gradually decreases via the $\beta$-decay thereafter (for a result of precise calculation, see Fig.~\ref{fig2}).  The extra neutron abundance at $t\gtrsim t_{\rm inj}$ is then approximately solved to be
\begin{eqnarray}
\Delta n(t)&\approx& \Delta n_{\rm inj} \left[\frac{a(t_{\rm inj})}{a(t)}\right]^3 \nonumber\\
&&\times \exp\left\{-\left[(n_{\rm H} \langle {\rm H}+n \rangle)_{t_{\rm inj}} +\tau_n^{-1} \right] (t-t_{\rm inj})\right\}, \nonumber\\
\label{eqb2}
\end{eqnarray}
where
$a(t)=1/(1+z)$ is the scale factor of universe with the redshift $z$, and
quantities with subscript $t_{\rm inj}$ represent values at time $t_{\rm inj}$.

The abundance of extra D is given by an integration of production rate via $^1$H($n,\gamma$)$^2$H.  Assuming an instantaneous production of extra D, its abundance at $t\gtrsim t_{\rm inj}$ is given by
\begin{eqnarray}
\Delta n_{\rm D}&=&\int_{t_{\rm inj}}^t n_{\rm H} \Delta n \langle {\rm H}+n \rangle~dt\nonumber\\
&\approx& (n_{\rm H} \langle {\rm H}+n \rangle)_{t_{\rm inj}}\int_{t_{\rm inj}}^t \Delta n~dt\nonumber\\
&=&\frac{(n_{\rm H} \langle {\rm H}+n \rangle)_{t_{\rm inj}}}{(n_{\rm H} \langle {\rm H}+n \rangle)_{t_{\rm inj}}+\tau_n^{-1}}\Delta n_{\rm inj}.
\label{eqb3}
\end{eqnarray}
The change in D abundance at the neutron injection is then given by
\begin{equation}
\left(\frac{\Delta {\rm D}}{\rm H}\right)_{t_{\rm inj}}
=\left(\frac{\Delta n_{\rm D}}{n_{\rm H}}\right)_{t_{\rm inj}}
\approx \frac{\langle {\rm H}+n \rangle_{t_{\rm inj}}}{(n_{\rm H} \langle {\rm H}+n \rangle)_{t_{\rm inj}}+\tau_n^{-1}}\Delta n_{\rm inj}.
\label{eqb4}
\end{equation}

\subsection{$^7$Be transformation}
When an abundance of extra neutron is much larger than that of thermal background neutron, the evolution of $^7$Be abundance is described by
\begin{equation}
\frac{d n_{^7{\rm Be}}}{dt}=-3H n_{^7{\rm Be}} -\Delta n n_{^7{\rm Be}}\langle ^7{\rm Be}+n \rangle.
\label{eqb5}
\end{equation}
Using Eq. (\ref{eqb2}), an approximate solution is obtained:
\begin{eqnarray}
n_{^7{\rm Be}}(t)&\hspace{-5pt}=&\hspace{-5pt}n_{^7{\rm Be}}(t_{\rm inj}) \left[\frac{a(t_{\rm inj})}{a(t)}\right]^3 \exp\left[-\int_{t_{\rm inj}}^t \Delta n \langle ^7{\rm Be}+n \rangle~dt\right]\nonumber\\
&\hspace{-5pt}\approx&\hspace{-5pt} n_{^7{\rm Be}}(t_{\rm inj}) \left[\frac{a(t_{\rm inj})}{a(t)}\right]^3\nonumber\\
&\hspace{-5pt}&\hspace{-5pt}\times \exp\left[-\frac{\langle ^7{\rm Be}+n \rangle_{t_{\rm inj}}}{(n_{\rm H} \langle {\rm H}+n \rangle)_{t_{\rm inj}}+\tau_n^{-1}}\Delta n_{\rm inj}\right].
\label{eqb6}
\end{eqnarray}

The change in $^7$Be abundance is then
\begin{eqnarray}
\left(\frac{-\Delta ^7{\rm Be}}{\rm H}\right)_{t_{\rm inj}} &\hspace{-10pt}=&\hspace{-5pt}\left(\frac{^7{\rm Be}}{\rm H}\right)_{t_{\rm inj}}\nonumber\\
&\hspace{-25pt}\times&\hspace{-15pt} \left\{1-\exp\left[-\frac{\langle ^7{\rm Be}+n \rangle_{t_{\rm inj}}}{(n_{\rm H} \langle {\rm H}+n \rangle)_{t_{\rm inj}}+\tau_n^{-1}}\Delta n_{\rm inj}\right]\right\}.\nonumber\\
\label{eqb7}
\end{eqnarray}

\subsection{D survival}
The survival fraction of D produced by the extra neutrons is estimated using a simplified rate equation for D, i.e.,
\begin{equation}
\frac{d(\rm D/H)}{dt}=-\frac{2({\rm D/H})^2}{2}~n_{\rm H} \langle {\rm D}+{\rm D} \rangle,
\label{eqb8}
\end{equation}
where
the factors of two in numerator and denominator in the RHS are for the number of D lost in one reaction, and for avoiding a double counting of initial state D nuclei, respectively.  We approximately take the hydrogen number density to be constant.  The survival fraction is then given by
\begin{eqnarray}
P_{\rm sur}({\rm D})&\equiv&\frac{{\rm D/H}(t)}{\left({\rm D/H}\right)_{t_{\rm inj}}} \nonumber\\
&\approx&\left[1+\left(\frac{\rm D}{\rm H}\right)_{t_{\rm inj}} \int_{t_{\rm inj}}^t n_{\rm H} \langle {\rm D}+{\rm D} \rangle~dt\right]^{-1}.~~~~~
\label{eqb9}
\end{eqnarray}
In the above equation, $($D/H$)_{t_{\rm inj}}$ is given by the sum of value in the SBBN model, i.e., $($D/H$)_{\rm SBBN}$, plus $(\Delta$D/H$)_{t_{\rm inj}}$ [Eq. (\ref{eqb4})].

\subsection{$^7$Li destruction}
The destruction fraction of $^7$Li produced via the conversion of $^7$Be by neutron capture is roughly estimated taking account of only the instantaneous proton burning of $^7$Li (see Appendix \ref{app1}).  The evolution of $^7$Li abundance is described by
\begin{equation}
\frac{d n_{^7{\rm Li}}}{dt}=-3H n_{^7{\rm Li}} -n_{\rm H} n_{^7{\rm Li}}\langle ^7{\rm Li}+p \rangle.
\label{eqb10}
\end{equation}
The destruction fraction of $^7$Li is then given by
\begin{eqnarray}
P_{\rm des}(^7{\rm Li})&\equiv&1-\frac{^7{\rm Li/H}(t)}{\left(^7{\rm Li/H}\right)_{t_{\rm inj}}}\nonumber \\
&=&1-\exp\left[-\int_{t_{\rm inj}}^t~n_{\rm H} \langle ^7{\rm Li}+p \rangle~dt\right].~~~
\label{eqb11}
\end{eqnarray}

\begin{figure}
\begin{center}
\includegraphics[width=8.0cm,clip]{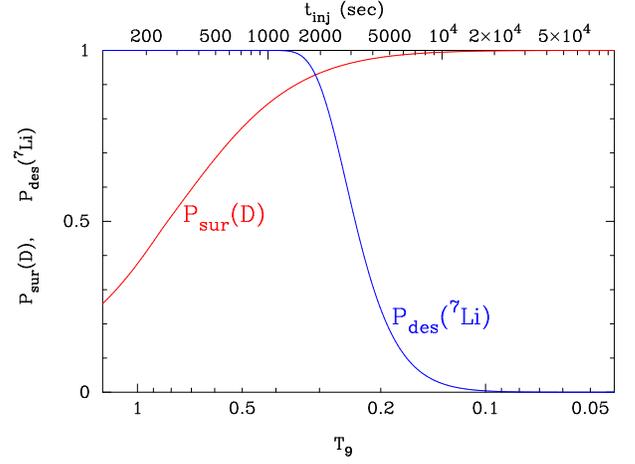}
\caption{Fraction of D which survives the destruction at collision with D [$P_{\rm sur}({\rm D})$], and fraction of $^7$Li which is destroyed by proton capture [$P_{\rm des}(^7{\rm Li})$] after their propagations through the universe from the time ($t_{\rm inj}$) or the temperature [$T_9\equiv T/(10^9~{\rm K})$] of their productions.  In drawing curves, approximate Eqs. (\ref{eqb9}) and (\ref{eqb11}) are used.\label{fig6}}
\end{center}
\end{figure}

Figure \ref{fig6} shows approximate values of $P_{\rm sur}({\rm D})$ and $P_{\rm des}(^7{\rm Li})$ calculated with Eqs. (\ref{eqb9}) and (\ref{eqb11}).  The increase of D abundance relative to that in the SBBN is not taken into account in the curve for $P_{\rm sur}({\rm D})$.  When the amount of extra neutron injection is larger, (D/H)$_{\rm inj}$ is larger, and resultingly the $P_{\rm sur}({\rm D})$ value decreases by the self destruction of D [Eq. (\ref{eqb9})].  A neutron injection at lower temperature triggers a production of D with a higher survival probability.   The $^7$Li nuclei produced via the conversion of $^7$Be has a smaller destruction probability at lower temperature.


In Fig. \ref{fig3}, thin dotted lines correspond to abundance ratios, i.e., D/H$=($D/H$)_{\rm SBBN}+(\Delta $D/H$)_{t_{\rm inj}}P_{\rm sur}({\rm D})$ [with Eqs. (\ref{eqb4}) and (\ref{eqb9})] and $^7$Li/H$=(^7$Li/H$)_{\rm SBBN}-(-\Delta ^7{\rm Be}/$H$)_{\rm inj}P_{\rm des}(^7{\rm Li})$ [with Eqs. (\ref{eqb7}) and (\ref{eqb11})], respectively.  In calculating the ratios, we read abundance evolution profiles in the SBBN model and used them.  The analytical result (dotted lines) are rather consistent with the results of full calculation (solid and dashed lines).

\begin{acknowledgments}
This work is supported by the National Research Foundation of Korea (Grant Nos. 2012R1A1A2041974, 2011-0015467, 2012M7A1A2055605).
\end{acknowledgments}



\end{document}